\documentclass[useAMS,usenatbib]{mn2e}
\usepackage{graphicx}
\begin{document}

\title[A spectroscopic study of the Globular Cluster M28 (NGC~6626)]{A spectroscopic study of the
  Globular Cluster M28 (NGC~6626)\footnote{Based on observations with the ESO Very Large Telescope at La Silla-Paranal Observatory, Chile (program ID 091.D-0535), and with ESO VISTA public survey VVV (program ID 179.B-2002)}}
\author[S. Villanova et al.]{S. Villanova$^{1}$\thanks{E-mail: svillanova@astro-udec.cl (SV)}, 
C. Moni Bidin$^{2}$, F. Mauro$^{1}$, C. Munoz$^{1}$, and L. Monaco$^{3}$\\
$^{1}$Departamento de Astronomia, Casilla 160, Universidad de Concepci{\'o}n, Chile\\
$^{2}$Instituto de Astronomia, Universidad Catolica del Norte, Av. Angamos 0610 Antofagasta Chile\\
$^{3}$Universidad Andres Bello, Departamento de Ciencias Fisicas, Republica 220, Santiago, Chile\\
}

\date{Accepted --. Received --; in original form --}

\pagerange{\pageref{firstpage}--\pageref{lastpage}} \pubyear{2016}

\maketitle

\label{firstpage}

\begin{abstract}
We present the abundance analysis for a sample of 17 red giant branch stars  in
the metal-poor globular cluster M28 based on high resolution spectra.
This is the first extensive spectroscopic study of this cluster. 
We derive abundances of O, Na, Mg, Al, Si, Ca, Ti, V, Cr, Mn, Fe, Co, Ni, Cu,
Zn, Y, Zr, Ba, La, Ce, and Eu. We find a metallicity of [Fe/H]=-1.29$\pm$0.01 and an
$\alpha$-enhancement of +0.34$\pm$0.01 (errors on the mean), 
typical of Halo Globular Clusters in this
metallicity regime. A large spread is observed in the abundances of light
elements O, Na, and Al. Mg also shows an anticorrelation with Al with a
significance of 3$\sigma$. The cluster shows a Na-O anticorrelation and a Na-Al correlation. This correlation is not linear
but ``segmented'' and that the stars are not
distributed continuously, but form at least 3 well
separated sub-populations. In this aspect M28 resembles NGC~2808 that was
found to host at least 5 sub-populations. The presence of a Mg-Al
anticorrelation favor massive AGB stars as the main polluters responsible for
the multiple-population phenomenon.
\end{abstract}


\section{Introduction}

Galactic Globular Clusters (GCs) are known to host star-to-star variations as
far as chemical abundances are concerned.  
More specifically, \citet{Ca09b} showed that all Galactic GCs studied up to now have at least a
spread (or anti-correlation) in the content of their light-elements O and
Na. In some cases also a Mg and Al spread is observed.
The only confirmed exception is Ruprecht 106, where \citet{Vi13} found that
stars share a homogenous chemical composition. 
This spread is due to the early evolution of each
cluster, that is initially formed by a first generation of stars that has the
same chemical composition of field stars at the same metallicity.
The subsequent generation of stars (Na-richer and O-poorer) are formed from gas polluted by
ejecta of evolved stars of the older generation (Na-poorer and O-richer). This is
the so called multiple-population phenomenon. 
These anomalies have also been observed in one of the old, massive extragalactic GCs in
Fornax \citep{Le06,La12} and in the Large Magellanic Cloud (LMC;
\citealt{Mu09}), but not in intermediate-age LMC clusters \citep{Mu14}.
This spectroscopic evidence has been interpreted as the signature of material
processed during H-burning by high temperature proton-capture reactions (such
as the Ne-Na and Mg-Al cycles). Several theoretical models have been proposed
to describe the formation and early evolution of GCs (e.g. \citealt{Da16}) 
The current explanation involves a self-enrichment scenario,
were subsequent generations of stars co-exist in globular clusters
that are formed from gas polluted by processed material
produced by massive stars \citep{Ca11}. Several
sources of polluters have been proposed: intermediate-mass asymptotic giant branch (AGB) stars \citep{Da16},
fast-rotating massive stars \citep{De07}, and massive binaries \citep{de09}. 
All these scenarios postulate that an important fraction of the first
generation has been lost in order to justify the relatively low fraction of
these stars that are observed nowaday compared with the objects of the
Na-richer and O-poorer generations.
A recent attractive alternative has been proposed by
\citet{Ba13} that implies only a single burst of star formation. They
postulate that the gas ejected from massive stars of the first (and the only)
generation concentrates in the center 
of the cluster and is acquired by low-mass stars via disk accretion, while
they are in the fully convective phase of the pre-main sequence. This scenario
has the advantage that does not require a huge star loss.
However it is not able to reproduce the Mg-Al anticorrelation that is
observed in some GCs like NGC~2808 \citep{Da16}.\\
In addition to the abundance spread of light elements, variations in heavier
elements have also been found in some massive GCs, such as $\omega$ Centauri 
\citep{Vi14}, M54 \citep{Ca10}, M22 \citep{Ma09}, Terzan5
\citep{Ma14} and NGC 2419 \citep{Co10}. However they are generally thought to
be the vestige of more massive primitive dwarf galaxies that merged with the
Galaxy.  

In this paper we present a spectroscopic study of the GC M28 (NGC~6626). This is an old
and metal-poor globular cluster that has received little attention, mostly due to
reddening problems and the strong field contamination since it is located
toward the Galactic Bulge slightly below the Galactic plane (l=7.80$^0$,
b=-5.58$^0$). M28 has a low metallicity of [Fe/H]=-1.32 and a high reddening
of E(B−V)=0.40 \citep{Ha96}.
The best photometry we could find in literature is that from \citet{Da96} that shows
a color-magnitude diagram (CMD) characteristic of metal-poor clusters with a 
horizontal branch (HB) that extends far to the blue. Its HB looks more like
that of a very metal poor GCs such as NGC~6397 ([Fe/H]=-2.0) or NGC~7078
([Fe/H]=-2.3) than that of a GC of the same metallicity such as NGC~1851
([Fe/H]=-1.2) or NGC~288 ([Fe/H]=-1.3) because M28 lacks completely a red HB.
From this we infer that M28 is significantly older than NGC~1851 or NGC~288
(11.0 and 11.5 Gyrs respectively, \citealt{Va13}) with an age
comparable to that of M12  ([Fe/H]=-1.3 and Age=13.0 Gyrs) based on the fact
that also M12 completely lacks a red HB \citep{Do14}. 
At only 2.7 kpc from the Galactic center \citep{Ha96}, M28 is one of the most
metal-poor GCs found in the inner Galaxy, and it was classified as a genuine
Bulge GCs by \citet{Bi15}. If it is really as old as
M12, M28 could be one of the oldest Bulge objects.

In section 2 we describe observation and data reduction and in section 3 the methodology we
used to obtain the chemical abundances and the associated errors. In section 4 we present our results
including a comparison with different environments (Galactic and
extragalactic). Finally in section 5 we give a summary of our findings.  

\begin{figure}
 \includegraphics[width=80mm,angle=270]{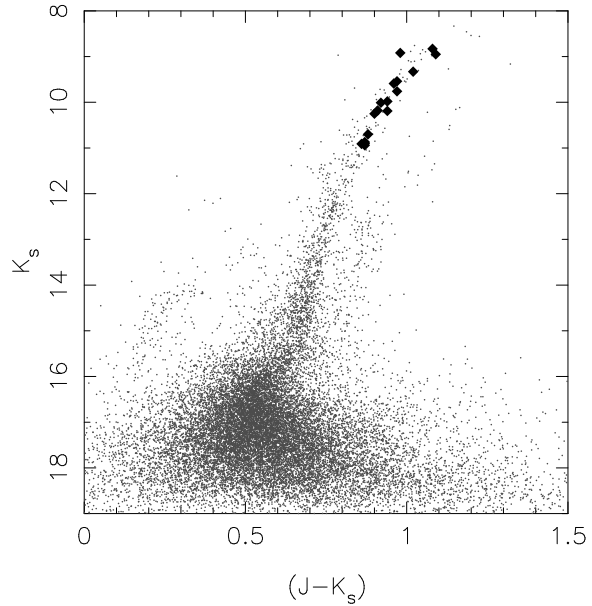}
 \caption{K$_{2MASS}$ vs. J-K$_{2MASS}$ CMD of M28. Target stars are indicated
 as filled black circles.}
 \label{f1}
\end{figure}

\begin{table*}
\caption{ID numbers, coordinates, VVV (J,H,K) magnitudes, heliocentric
  radial velocities and atmospheric parameters for the observed stars. Reported errors are errors on the mean.}            
\label{t1}      
\centering                          
\begin{tabular}{lccccccccccc}        
\hline
\hline                 
ID  & ID(2MASS) & \scriptsize{RA} & \scriptsize{DEC} & \scriptsize{J} & \scriptsize{H} & \scriptsize{K} & \scriptsize{RV$_{\rm H}$} & \scriptsize{T$_{eff}$} & \scriptsize{log(g)} & \scriptsize{v$_{t}$} & \scriptsize{[Fe/H]}\\    
    &           &   degree      s &    degrees       &  mag.          &  mag.          &   mag.         &    km/s                   &           K            &         dex         &       km/s           &       dex          \\    
\hline      
\hline
1236     & 18242731-2452349 & 276.113810 & -24.876362 & 11.811 & 11.129 & 10.944 &  2.6  & 4617 & 1.80 & 1.40 & -1.16\\
1239     & 18242792-2451160 & 276.116356 & -24.854464 & 11.748 & 11.045 & 10.874 &  0.0  & 4537 & 1.56 & 1.32 & -1.20\\
1246$_1$ & 18242914-2452024 & 276.121430 & -24.867336 & 10.935 & 10.197 & 10.014 & 18.8  & 4218 & 0.94 & 1.39 & -1.30\\
1246$_2$ & 18242914-2452024 & 276.121430 & -24.867336 & 10.935 & 10.197 & 10.014 & 19.9  & 4252 & 1.00 & 1.42 & -1.32\\
1280     & 18243130-2452471 & 276.130424 & -24.879751 & 10.722 &  9.969 &  9.756 & 18.0  & 4219 & 0.99 & 1.49 & -1.38\\
1282     & 18243140-2451136 & 276.130842 & -24.853786 & 10.042 &  9.125 &  8.951 &  8.8  & 3983 & 0.44 & 1.56 & -1.38\\
1291     & 18243191-2453293 & 276.132983 & -24.891479 & 10.351 &  9.533 &  9.335 &  7.0  & 4091 & 0.64 & 1.48 & -1.32\\
1293$_1$ & 18243208-2452567 & 276.133679 & -24.882437 &  9.910 &  9.053 &  8.827 & 26.8  & 3945 & 0.49 & 1.55 & -1.38\\
1293$_2$ & 18243208-2452567 & 276.133679 & -24.882437 &  9.910 &  9.053 &  8.827 & 28.0  & 3974 & 0.53 & 1.54 & -1.34\\
1295     & 18243227-2452471 & 276.134463 & -24.879768 & 10.510 &  9.746 &  9.545 & 12.6  & 4180 & 0.84 & 1.47 & -1.32\\
1315     & 18243318-2451079 & 276.138277 & -24.852196 & 11.125 & 10.372 & 10.186 &  1.0  & 4297 & 1.04 & 1.43 & -1.27\\
1328     & 18243398-2451227 & 276.141588 & -24.856312 & 10.924 & 10.181 &  9.983 & 11.8  & 4301 & 1.14 & 1.42 & -1.23\\
1330     & 18243399-2450583 & 276.141626 & -24.849554 & 11.581 & 10.864 & 10.701 & 15.1  & 4426 & 1.27 & 1.28 & -1.28\\
1343     & 18243460-2452406 & 276.144202 & -24.877949 & 11.774 & 11.095 & 10.905 & 24.7  & 4571 & 1.68 & 1.35 & -1.20\\
1364     & 18243590-2452090 & 276.149619 & -24.869192 & 11.768 & 11.089 & 10.910 &  8.9  & 4508 & 1.46 & 1.36 & -1.30\\
1367     & 18243627-2452221 & 276.151166 & -24.872833 & 11.086 & 10.339 & 10.178 & 19.3  & 4323 & 1.22 & 1.40 & -1.34\\
1378$_1$ & 18243700-2451179 & 276.154185 & -24.854994 & 10.551 &  9.763 &  9.587 & -2.0  & 4161 & 0.85 & 1.48 & -1.26\\
1378$_2$ & 18243700-2451179 & 276.154185 & -24.854994 & 10.551 &  9.763 &  9.587 & -1.4  & 4107 & 0.64 & 1.42 & -1.32\\
1380$_1$ & 18243713-2452257 & 276.154724 & -24.873831 &  9.900 &  9.040 &  8.919 & 18.6  & 4098 & 0.80 & 1.53 & -1.24\\
1380$_2$ & 18243713-2452257 & 276.154724 & -24.873831 &  9.900 &  9.040 &  8.919 & 19.4  & 4052 & 0.54 & 1.46 & -1.28\\
1402     & 18243900-2451077 & 276.162541 & -24.852152 & 11.146 & 10.397 & 10.246 & 20.3  & 4391 & 1.32 & 1.46 & -1.27\\
\hline                                    
Cluster  &                  &            &            &        &        &        & 13.3  &      &      &      & -1.29\\
Error    &                  &            &            &        &        &        &  2.0  &      &      &      &  0.01\\
\hline
\end{tabular}
\end{table*}

\begin{table*}
\caption{Chemical abundances of our stars (Part 1). The abundance for Ti is the
    mean of those obtained from the neutral and singly ionized
    species. Reported errors are errors on the mean.}            
\label{t2}      
\centering                          
\begin{tabular}{lcccccccccc}        
\hline\hline                 
ID & \scriptsize{[OI/Fe]} & \scriptsize{[NaI/Fe]} & \scriptsize{[MgI/Fe]} &\scriptsize{[AlI/Fe]} & \scriptsize{[SiI/Fe]} & \scriptsize{[CaI/Fe]} & \scriptsize{[Ti/Fe]} & \scriptsize{[VI/Fe]} & \scriptsize{[CrI/Fe]} & \scriptsize{[MnI/Fe]}\\    
\hline      
1236      & -0.58 &  0.41 & 0.43 & 1.01 & 0.28 & 0.39 & 0.36 & 0.45 &  0.02 & -0.47 \\
1239      & -0.51 &  0.44 & 0.50 & 1.06 & 0.32 & 0.31 & 0.30 & 0.24 & -0.01 & -0.42 \\
1246$_1$  &  0.05 &  0.43 & 0.46 & 0.48 & 0.35 & 0.33 & 0.34 & 0.34 & -0.03 & -0.49 \\
1246$_2$  & -0.08 &  0.39 & 0.55 & 0.46 & 0.31 & 0.35 & 0.31 & 0.27 & -0.09 & -0.39 \\
1280      &  0.35 & -0.06 & 0.52 & 0.30 & 0.34 & 0.41 & 0.34 & 0.35 &  0.07 & -0.34 \\
1282      &  0.29 &  0.00 & 0.49 & 0.17 & 0.35 & 0.40 & 0.30 & 0.33 & -0.04 & -0.36 \\
1291      & -0.72 &  0.58 & 0.44 & 1.07 & 0.37 & 0.39 & 0.29 & 0.35 &  0.03 & -0.34 \\
1293$_1$  &  0.23 & -0.05 & 0.44 & 0.24 & 0.38 & 0.35 & 0.26 & 0.29 &  0.07 & -0.39 \\
1293$_2$  &  0.32 & -0.02 & 0.46 & 0.20 & 0.31 & 0.35 & 0.29 & 0.42 & -0.05 & -0.37 \\
1295      & -0.79 &  0.72 & 0.44 & 1.02 & 0.37 & 0.40 & 0.31 & 0.35 &  0.00 & -0.35 \\
1315      & -0.17 &  0.38 & 0.43 & 0.73 & 0.40 & 0.42 & 0.32 & 0.31 &  0.04 & -0.39 \\
1328      & -0.61 &  0.46 & 0.36 & 1.10 & 0.38 & 0.36 & 0.32 & 0.33 & -0.01 & -0.40 \\
1330      & -0.55 &  0.54 & 0.45 & 1.05 & 0.43 & 0.42 & 0.36 & 0.22 & -0.08 & -0.38 \\
1343      &  0.17 &  0.17 & 0.51 & 0.32 & 0.24 & 0.30 & 0.33 & 0.43 &  0.04 & -0.42 \\
1364      & -0.46 &  0.44 & 0.49 & 0.93 & 0.31 & 0.40 & 0.36 & 0.29 & -0.03 & -0.35 \\
1367      & -0.04 &  0.42 & 0.50 & 0.42 & 0.40 & 0.40 & 0.34 & 0.28 & -0.19 & -0.39 \\
1378$_1$  &  0.28 & -0.07 & 0.45 & 0.11 & 0.28 & 0.39 & 0.32 & 0.37 & -0.10 & -0.33 \\
1378$_2$  &  0.20 &  0.02 & 0.53 & 0.15 & 0.34 & 0.38 & 0.28 & 0.21 & -0.08 & -0.39 \\
1380$_1$  &  0.30 & -0.10 & 0.50 & 0.22 & 0.23 & 0.36 & 0.31 & 0.44 & -0.02 & -0.34 \\
1380$_2$  &  0.20 & -0.04 & 0.49 & 0.16 & 0.33 & 0.35 & 0.27 & 0.37 &  0.00 & -0.37 \\
1402      & -0.44 &  0.55 & 0.47 & 1.03 & 0.33 & 0.37 & 0.34 & 0.44 & -0.02 & -0.36 \\
\hline        
Cluster   & -0.12 &  0.27 & 0.47 & 0.58 & 0.34 & 0.37 & 0.32 & 0.34 & -0.02 & -0.38 \\
Error     &  0.09 &  0.06 & 0.01 & 0.08 & 0.01 & 0.01 & 0.01 & 0.02 &  0.01 &  0.01 \\
\hline
\end{tabular}
\end{table*}

\begin{table*}
\caption{Chemical abundances of our stars (Part 2). Reported errors are errors on the mean.}            
\label{t3}      
\centering                          
\begin{tabular}{lcccccccccc}        
\hline\hline                 
ID & \scriptsize{[CoI/Fe]} & \scriptsize{[NiI/Fe]} & \scriptsize{[CuI/Fe]} & \scriptsize{[ZnI/Fe]} & \scriptsize{[YII/Fe]} & \scriptsize{[ZrII/Fe]} & \scriptsize{[BaII/Fe]} & \scriptsize{[LaII/Fe]} & \scriptsize{[CeII/Fe]} & \scriptsize{[EuII/Fe]} \\    
\hline      
1236      &  0.13 & -0.02 &    -  &  0.09 & 0.12 & 0.25 & 0.35 & -0.33 &  0.12 & 0.39\\
1239      &  0.14 & -0.01 & -0.29 &  0.19 & 0.09 & 0.30 & 0.34 & -0.18 & -0.05 & 0.41\\
1246$_1$  &  0.04 &  0.02 &   -   &   -   &  -   & 0.12 &  -   &   -   &   -   & 0.53\\
1246$_2$  &  0.09 & -0.04 & -0.17 &   -   & 0.17 & 0.24 & 0.35 & -0.07 & -0.01 & 0.47\\
1280      &  0.11 & -0.04 & -0.25 &  0.23 & 0.16 & 0.34 & 0.24 &  0.13 & -0.33 & 0.40\\
1282      &  0.10 & -0.05 & -0.20 &   -   & 0.16 & 0.29 & 0.20 &  0.06 & -0.28 & 0.32\\
1291      &  0.11 & -0.03 & -0.25 &  0.32 & 0.24 & 0.27 & 0.44 &  0.14 & -0.13 & 0.36\\
1293$_1$  &  0.08 &  0.00 & -0.11 & -0.04 & 0.07 & 0.26 & 0.20 &   -   & -0.07 & 0.39\\
1293$_2$  &  0.08 & -0.03 & -0.14 &  0.02 & 0.07 & 0.26 & 0.35 &  0.03 & -0.19 & 0.37\\
1295      &  0.10 & -0.05 & -0.24 &  0.26 & 0.27 & 0.26 & 0.41 & -0.06 & -0.14 & 0.36\\
1315      &  0.08 & -0.07 & -0.17 &  0.21 & 0.14 & 0.23 & 0.22 & -0.36 & -0.11 & 0.37\\
1328      &  0.08 & -0.06 & -0.34 &  0.31 & 0.21 & 0.22 & 0.27 & -0.09 & -0.21 & 0.13\\
1330      & -0.07 & -0.03 & -0.25 &  0.26 & 0.14 & 0.25 & 0.25 & -0.25 & -0.27 & 0.28\\
1343      &  0.08 & -0.01 &   -   &   -   & 0.01 & 0.26 & 0.20 &  0.01 & -0.19 & 0.29\\
1364      &  0.12 & -0.01 &   -   &  0.20 & 0.12 & 0.26 & 0.30 &  0.07 & -0.39 & 0.31\\
1367      &  0.03 &  0.03 & -0.29 &  0.04 & 0.22 & 0.28 & 0.23 &  0.17 & -0.15 & 0.46\\
1378$_1$  &  0.15 & -0.04 & -0.25 &  0.13 &  -   & 0.22 & 0.34 &   -   & -0.26 & 0.35\\
1378$_2$  &  0.09 & -0.10 & -0.22 &  0.35 & 0.11 & 0.24 & 0.26 & -0.15 & -0.27 & 0.24\\
1380$_1$  &  0.05 & -0.03 & -0.27 &  0.09 & 0.13 & 0.32 & 0.34 &  0.06 & -0.04 & 0.36\\
1380$_2$  &  0.08 & -0.05 & -0.19 &  0.23 & 0.15 & 0.25 & 0.36 &  0.02 & -0.20 & 0.35\\
1402      &  0.09 & -0.03 & -0.13 &  0.11 & 0.15 & 0.30 & 0.38 &  0.18 & -0.23 & 0.26\\
\hline    
Cluster   &  0.08 & -0.03 & -0.22 &  0.18 & 0.14 & 0.26 & 0.30 & -0.03 & -0.17 & 0.35\\
Error     &  0.01 &  0.01 &  0.02 &  0.03 & 0.01 & 0.01 & 0.02 &  0.04 &  0.03 & 0.02\\
\hline
\end{tabular}
\end{table*}

\begin{table*}
\caption{Estimated errors on abundances due to errors on atmospheric
parameters and to spectral noise for star \#1367 (column 2 to 6). Column 7
gives the total error calculated as the root squared of the sum of the squared
of columns 2 to 6. This total error must be compared with the total error as
obtained from the 4 repeated stars (column 8) and with the observed
dispersion (RMS) of the data with its error (column 9). The last column gives the significance of the difference between
the total error for star \#1367 and the observed dispersion, in units of $\sigma$.}
\label{t4}      
\centering                          
\begin{tabular}{lccccccccc}        
\hline\hline  
ID & $\Delta$T$_{\rm eff}$=40 K  & $\Delta$log(g)=0.12 & $\Delta$v$_{\rm t}$=0.04 km/s
& $\Delta$[Fe/H]=0.03 & S/N & $\Delta_{\rm tot}$ & $\Delta_{\rm tot}(obs)$  &$RMS_{\rm obs}$ &
Significance ($\sigma$)\\
\hline
$\Delta$([O/Fe])  & 0.02 & 0.06 & 0.01 & 0.01 & 0.05 & 0.08 & 0.07 & 0.40$\pm$0.06 & 5.3\\
$\Delta$([Na/Fe]) & 0.02 & 0.02 & 0.01 & 0.00 & 0.03 & 0.04 & 0.04 & 0.27$\pm$0.06 & 3.8\\
$\Delta$([Mg/Fe]) & 0.01 & 0.01 & 0.01 & 0.00 & 0.03 & 0.04 & 0.03 & 0.04$\pm$0.01 & 0.0\\
$\Delta$([Al/Fe]) & 0.01 & 0.02 & 0.01 & 0.01 & 0.04 & 0.05 & 0.03 & 0.39$\pm$0.06 & 5.7\\
$\Delta$([Si/Fe]) & 0.03 & 0.02 & 0.01 & 0.01 & 0.03 & 0.05 & 0.06 & 0.05$\pm$0.01 & 0.0\\
$\Delta$([Ca/Fe]) & 0.02 & 0.02 & 0.00 & 0.01 & 0.04 & 0.05 & 0.01 & 0.04$\pm$0.01 & 1.0\\
$\Delta$([Ti/Fe]) & 0.03 & 0.02 & 0.00 & 0.01 & 0.03 & 0.05 & 0.02 & 0.03$\pm$0.01 & 2.0\\
$\Delta$([V/Fe])  & 0.04 & 0.02 & 0.00 & 0.01 & 0.05 & 0.07 & 0.09 & 0.07$\pm$0.01 & 0.0\\
$\Delta$([Cr/Fe]) & 0.03 & 0.02 & 0.00 & 0.01 & 0.04 & 0.06 & 0.05 & 0.06$\pm$0.01 & 0.0\\
$\Delta$([Mn/Fe]) & 0.03 & 0.01 & 0.01 & 0.00 & 0.02 & 0.04 & 0.05 & 0.04$\pm$0.01 & 0.0\\
$\Delta$([Fe/H])  & 0.05 & 0.02 & 0.01 & 0.01 & 0.01 & 0.06 & 0.03 & 0.06$\pm$0.01 & 0.0\\
$\Delta$([Co/Fe]) & 0.02 & 0.01 & 0.01 & 0.00 & 0.06 & 0.07 & 0.03 & 0.05$\pm$0.01 & 2.0\\
$\Delta$([Ni/Fe]) & 0.00 & 0.02 & 0.00 & 0.00 & 0.02 & 0.03 & 0.02 & 0.03$\pm$0.01 & 0.0\\
$\Delta$([Cu/Fe]) & 0.01 & 0.02 & 0.01 & 0.01 & 0.07 & 0.08 & 0.04 & 0.06$\pm$0.01 & 2.0\\
$\Delta$([Zn/Fe]) & 0.05 & 0.03 & 0.01 & 0.00 & 0.09 & 0.11 & 0.06 & 0.11$\pm$0.02 & 0.0\\
$\Delta$([Y/Fe])  & 0.03 & 0.05 & 0.01 & 0.01 & 0.03 & 0.08 & 0.01 & 0.06$\pm$0.01 & 2.0\\
$\Delta$([Zr/Fe]) & 0.03 & 0.04 & 0.01 & 0.01 & 0.03 & 0.06 & 0.06 & 0.04$\pm$0.01 & 2.0\\ 
$\Delta$([Ba/Fe]) & 0.03 & 0.02 & 0.03 & 0.00 & 0.03 & 0.06 & 0.08 & 0.07$\pm$0.01 & 1.0\\
$\Delta$([La/Fe]) & 0.02 & 0.06 & 0.00 & 0.01 & 0.08 & 0.10 &  -   & 0.16$\pm$0.03 & 2.0\\
$\Delta$([Ce/Fe]) & 0.03 & 0.05 & 0.01 & 0.01 & 0.06 & 0.09 & 0.06 & 0.12$\pm$0.02 & 1.5\\
$\Delta$([Eu/Fe]) & 0.03 & 0.05 & 0.01 & 0.01 & 0.05 & 0.08 & 0.03 & 0.09$\pm$0.01 & 1.0\\
\hline                                   
\end{tabular}
\end{table*}

\section{Observations and data reduction}

Our dataset consists of high resolution spectra collected at the FLAMES@UVES
spectrograph mounted at the VLT-UT2 telescope. Targets were selected from the
infrared photometry collected with the Vista Variables in the Via Lactea
survey \citep[VVV,][]{Mi10,Sa12}. 
The PSF photometry was obtained
with the VVV-SkZ\_pipeline \citep{Ma13} on the publicly available
pre-processed frames, and calibrated in the astrometric and photometric 2MASS
system \citep{Sk06} as detailed in \citet{Mo11} and
\citet{Ch12}. A total of 17 stars were selected along the upper red giant
cluster sequence with magnitude between $K_s$=8.5 and $K_s$=11. The position
of the targets in the cluster VVV color-magnitude diagram (CMD) is shown in
Fig~\ref{f1}. These stars were observed with four fiber configurations of
FLAMES@UVES. We used the 580nm set-up, that gives a spectral coverage between
4800 and 6800~\AA\ with a resolution of R=47000. The signal-to-noise (S/N) was
between 50 and 70 at 6000~\AA. Four targets were observed twice, in two
different fiber configurations, ito perform a proper error analysis.

Data were reduced using the dedicated pipeline \footnote{see http://www.eso.org/sci/software/pipelines/}.
Data reduction includes bias subtraction, flat-field correction, wavelength calibration,
sky subtraction, and spectral rectification.

Radial velocities were measured by the {\it fxcor} package in IRAF \footnote{IRAF is distributed by the National
Optical Astronomy Observatory, which is operated by the Association of
Universities for Research in Astronomy, Inc., under cooperative agreement
with the National Science Foundation.}, using a synthetic spectrum as a template. 
The mean radial velocity we obtained is 13.3$\pm$2.0 km/s. \citet{Ha96}
gives 17$\pm$1 instead. The agreement is good if we consider that the cluster 
has a large velocity dispersion as shown by the R.M.S. of our stars that is
9.4$\pm$1.5 km/s. Thare are not clear outliers in the radial velocity
distribution. If we consider also that all our targets have the same
[Fe/H] content within errors (see Section 4) and that all lie along the Red
Giant sequence in the cluster CMD, we conclude that they are all
cluster member.
Table~\ref{t1} lists the basic parameters of the retained stars:
ID (stars observed twice are indicated by an underscored number), 2MASS
ID, J2000.0 coordinates (RA \& DEC in degrees), VVV $J,H,K_s$ magnitudes,
heliocentric radial velocity RV$_{\rm H}$ (km/s),
T$_{\rm {eff}}$ (K), log(g), micro-turbulence velocity v$_{\rm t}$ (km/s),
and [Fe/H] abundances. The determination of the atmospheric parameters
and abundances is discussed in the next section. In this table we report also
the cluster mean radial velocity and mean [Fe/H] abundance with their errors
(errors on the mean).
In Fig.~\ref{f1} we report, on the top of the M28 color magnitude diagram
(CMD), our targets as black filled points. 
We warn the reader that the four targets observed twice have two points
for each plot of this paper. We preferred to keep the measurements (abundances
and atmospheric parameters) of each of the two spectra per target separated
both in the text, in the tables and in the plots in order to allow a direct check of the
internal errors of our analysis. When required the two measurements per stars
will be indicated by different symbols in the plots and by the underscores 1
and 2 in the text and in the tables.

\section{Abundance analysis}

Initial atmospheric parameters were obtained in the following way.  
First, T$_{\rm eff}$ was derived from the J-K color using the relation of
\citet{Ra05}. The reddening we adopted (E(B-V)=0.40) was
obtained from \citet[2010 edition]{Ha96}. 
Surface gravities (log(g)) were obtained from the canonical equation:
$$ \log\left(\frac{g}{g_{\odot}}\right) =
         \log\left(\frac{M}{M_{\odot}}\right)
         + 4 \log\left(\frac{T_{\rm{eff}}}{T_{\odot}}\right)
         - \log\left(\frac{L}{L_{\odot}}\right). $$
where the mass M was assumed to be 0.8 M$_{\odot}$, and the
luminosity L/L$_{\odot}$ was obtained from the absolute magnitude M$_{\rm V}$
assuming an apparent distance modulus of (m-M)$_{\rm V}$=15.55 \citep{Ha96}. The
bolometric correction (BC) was derived by adopting the relation 
BC-T$_{\rm eff}$ from \citet{Al99}.
Finally, micro-turbulence velocity (v$_{\rm t}$) was obtained from the
relation of \citet{Ma08}.
Atmospheric models were calculated using ATLAS9 code \citep{Ku70}
assuming our estimations of T$_{\rm eff}$, log(g), and v$_{\rm t}$, and the
[Fe/H] value from \citet{Ha96}([Fe/H]=-1.32).\\ 
Then T$_{\rm eff}$, log(g), and v$_{\rm t}$ were re-adjusted and new 
atmospheric models calculated in an interactive way in order to remove trends 
in excitation potential and reduced equivalent width (EQW) versus abundance for T$_{\rm eff}$ and v$_{\rm t}$, respectively, 
and to satisfy the ionization equilibrium for log(g). 140$\div$150 FeI lines and
 12$\div$14 FeII lines (depending on the S/N of the spectrum) were used for the
latter purpose. The [Fe/H] value of the model was changed at each 
iteration according to the output of the abundance analysis. 
The Local Thermodynamic Equilibrium (LTE) program MOOG \citep{Sn73} was used
for the abundance analysis.

SiI, CaI, TiI, TiII, CrI, FeI, FeII, and NiI abundances were estimated
using the EQW method. For this purpose we measured EQW using the automatic
program DAOSPEC \citep{St08} \footnote{DAOSPEC is freely distributed by http://www.cadc-ccda.hia-iha.nrc-cnrc.gc.ca/en/community/STETSON/daospec/}. 
OI, NaI, MgI, AlI, VI, MnI, CoI, CuI, ZnI, YII, ZrII, BaII, LaII, CeII,
and EuII abundances were obtained using the spectro-synthesis method. For this
purpose 5 synthetic spectra were generated for each line with 0.25 dex
abundance step and compared with the observed spectrum. The 
line-list and the methodology we used are the same used in previous papers (e.g. \citealt{Vi13}),
so we refer to those articles for a detailed discussion about this point. 
Here we just underline that we took hyperfine splitting into account for Ba
as in our previous studies. This is particularly important because Ba
lines are very strong even in metal-poor stars and hyperfine splitting help to
remove the line-core saturation producing a change in the final abundance as
estimated by the spectro-synthesis method up to 0.1 dex. Also other
odd-elements like V, Mn, Co, Cu, Y, and Eu or elements that have odd-isotops
like La and Ce have an hyperfine splitting, but their lines are weak and the
line-core saturation is not at work. So hyperfine splitting corrections are negligible.

We also added the MgI line at 5528 \AA. Parameters for this line were taken
from SPECTRUM v2.76\footnote{http://www.appstate.edu/~grayro/spectrum/spectrum.html} linelist \citep{Gr94}.
The abundances we obtained are reported in Tab.~\ref{t2} and Tab.~\ref{t3} together with the mean values for
the cluster and the error on the mean. For Ti we reported the mean values of TiI and TiII abundances.
Na is an element affected by NLTE effets. For this reason we looked in
the INSPEC \footnote{version 1.0 (http://inspect.coolstars19.com/index.php?n=Main.HomePage)} 
database for suitable NLTE corrections. We found the they are very small
($\sim$-0.05 dex) with no significant variation (less then 0.02 dex) in our
temperature range. For this reason we decided not to apply them to our Na abundances.

\begin{figure}
 \includegraphics[width=80mm]{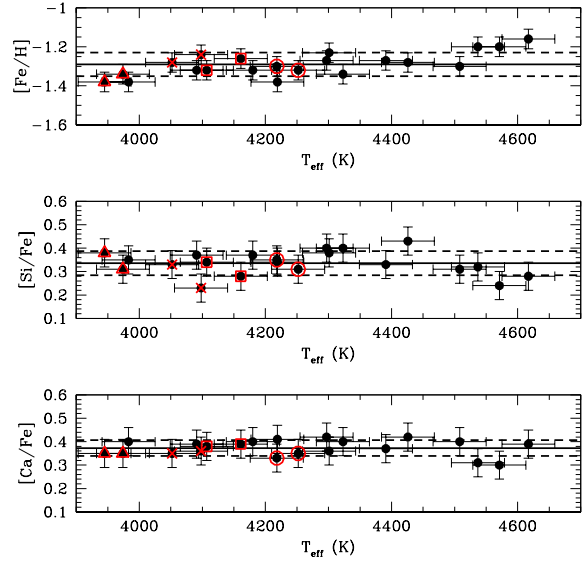}
 \caption{[Fe/H], [Si/Fe], and [Ca/Fe] vs. T$_{\rm eff}$ for our
   targets. The 4 stars observed twice are indicated with 4 different red open symbols.}
 \label{f2a}
\end{figure}

\begin{figure}
 \includegraphics[width=80mm]{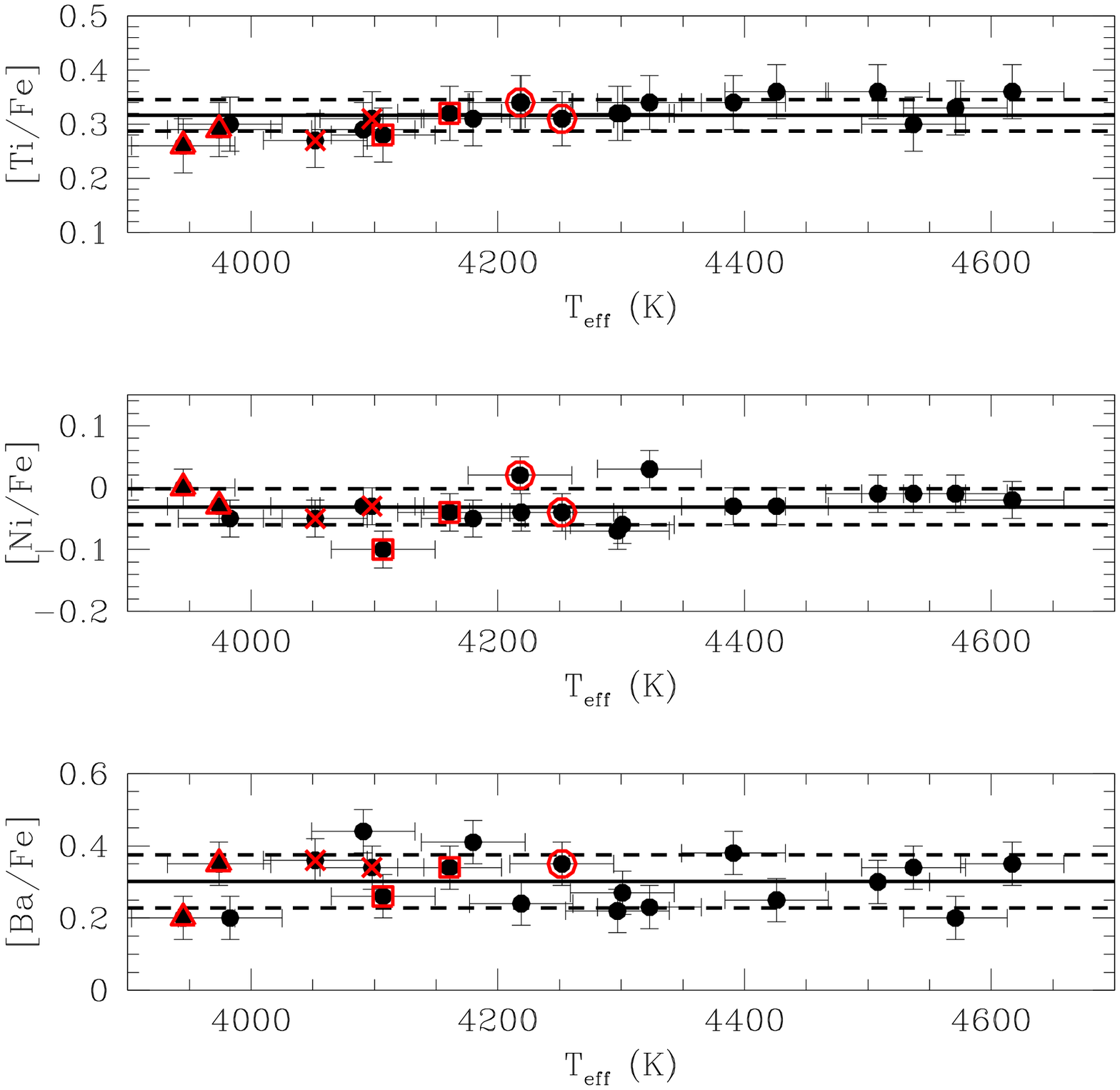}
 \caption{[Ti/Fe], [Ni/Fe], and [Ba/Fe] vs. T$_{\rm eff}$ for our
   targets. The 4 stars observed twice are indicated with 4 different red
   open symbols.}

 \label{f2aa}
\end{figure}

\begin{figure}
 \includegraphics[width=80mm]{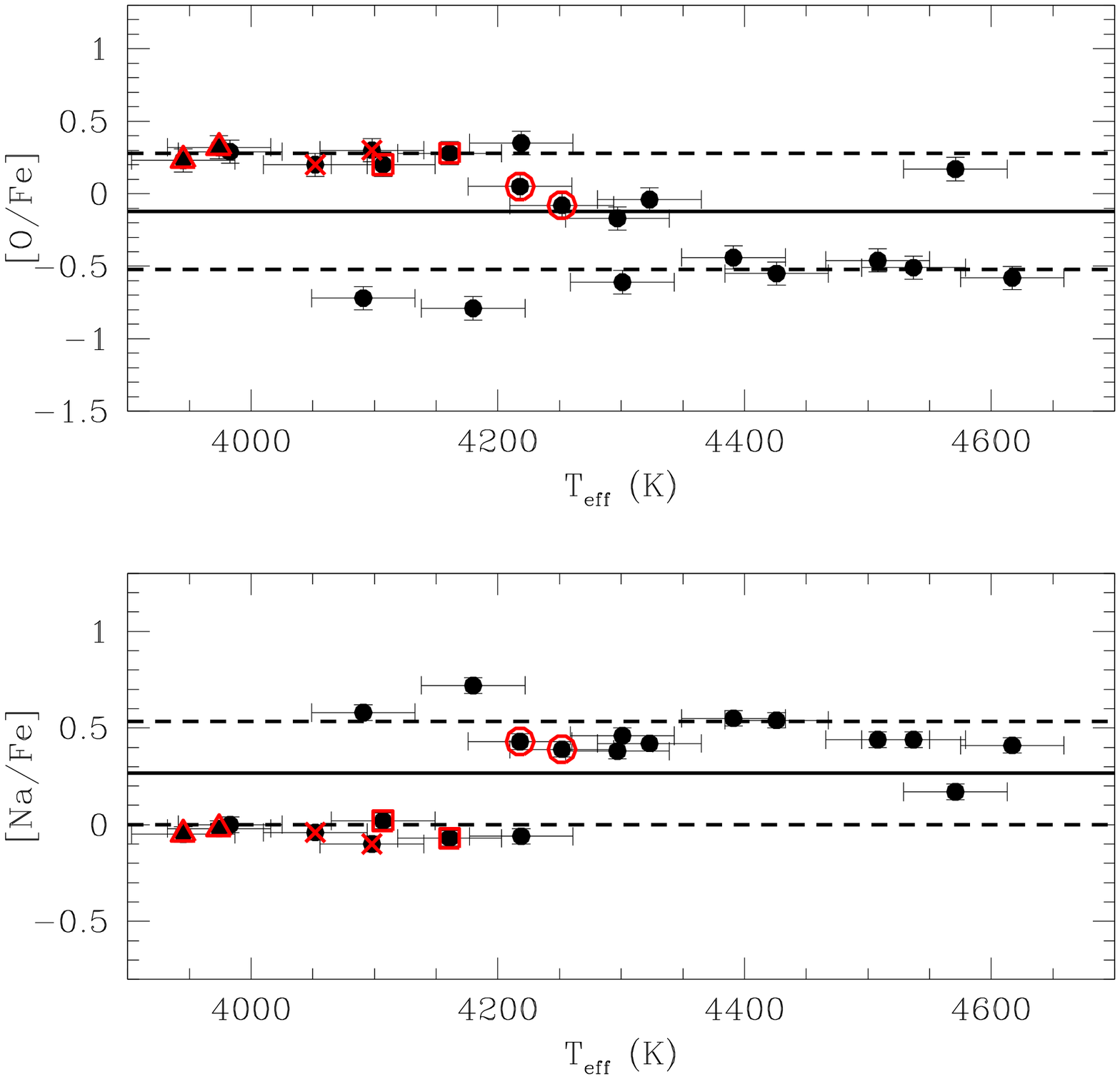}
 \caption{[O/Fe] and [Na/Fe] vs. T$_{\rm eff}$ for our targets. The 4
     stars observed twice are indicated with 4 different red open symbols.}
 \label{f2b}
\end{figure}

\begin{figure}
 \includegraphics[width=80mm]{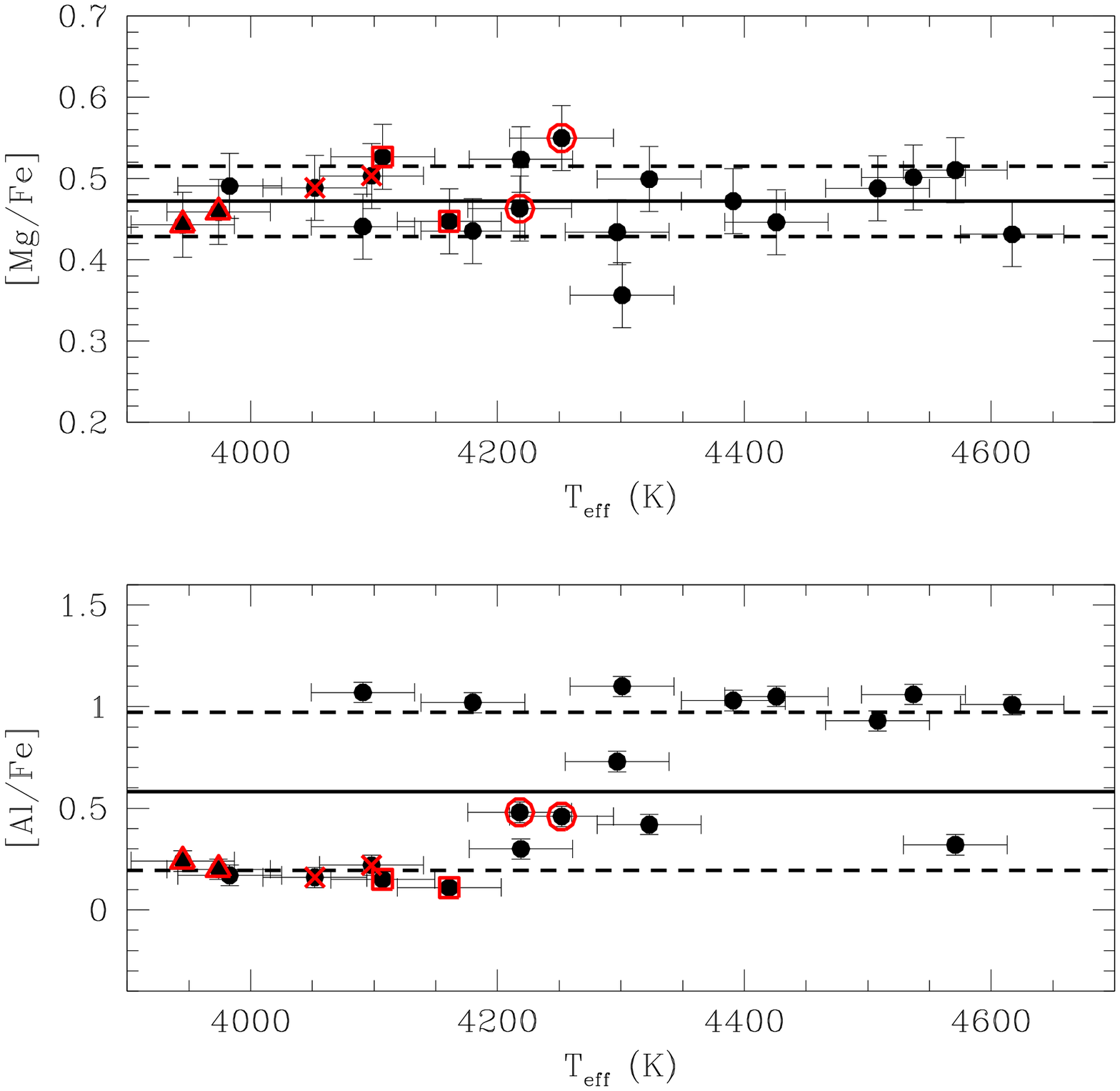}
 \caption{[Mg/Fe] and [Al/Fe] vs. T$_{\rm eff}$ for our targets. The  4
     stars observed twice are indicated with 4 different red open symbols.}
 \label{f2c}
\end{figure}

\begin{figure}
 \includegraphics[width=80mm]{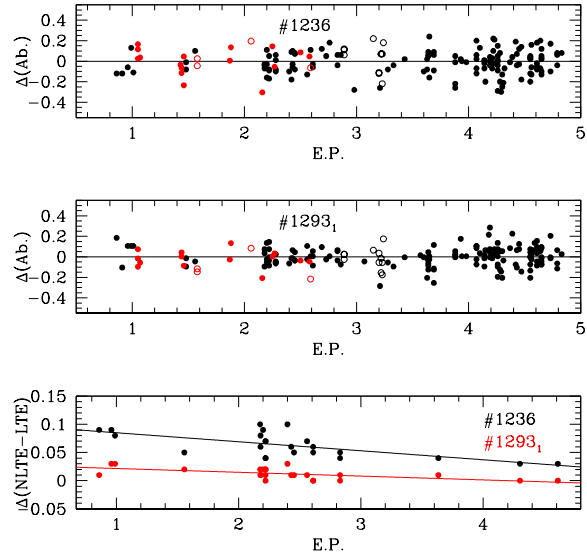}
 \caption{Upper panel: FeI (filled black circles), FeII (open black circles),
   TiI (filled red circles), and TiII (open red circles) differential
   abundances for stars \#1236. Middle panel: FeI (filled black circles), FeII
   (open black circles), TiI (filled red circles), and TiII (open red circles)
   differential abundances for stars \#1293$_1$.
    Lower panel: NLTE correction as a function of E.P. for iron lines for star \#1236
   (black points, T$_{\rm eff}$=4617 K) and \#1293$_1$ (red points, 
   T$_{\rm eff}$$\sim$3950 K). Linear fits are indicated as continuos lines.}
 \label{f2d}
\end{figure}

\begin{figure}
 \includegraphics[width=80mm]{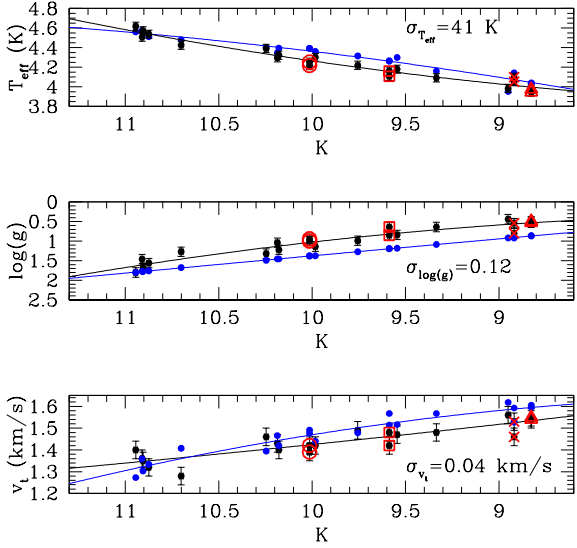}
 \caption{ T$_{\rm eff}$, log(g), and v$_{\rm t}$ as a function of K
   magnitude for our targets. The 4 stars observed twice are indicated
   with 4 different red open symbols. Temperature values are given divided by 1000.}
 \label{f3}
\end{figure}

As a cross check of our abundance analysis, we plot in Fig.~\ref{f2a}, Fig.~\ref{f2aa},
Fig.~\ref{f2b} and Fig.~\ref{f2c}  [Fe/H], [Si/Fe], and [Ca/Fe] vs. T$_{eff}$,
[Ti/Fe], [Ni/Fe], and [Ba/Fe] vs. T$_{eff}$, [O/Fe] and [Na/Fe] vs.  T$_{eff}$, and [Mg/Fe] and [Al/Fe] vs. T$_{eff}$
respectively, for the entire sample. The temperature range covered by our stars is about 600
K. We plot also the mean abundance for each element and the $\pm$1$\sigma$
error. 

[Fe/H] shows a trend as a funcion of temperature
with a significance on the slope of about 4 $\sigma$, with warmer stars
beeing more metal rich. In order to investigate this, first of all we plot
in Fig.~\ref{f2d} (upper and middle panels) the line by line FeI and FeII differential
abundances as a function of E.P. for stars \#1236 and \#1293$_1$. These stars
were  selected to be at the extremes of the T$_{eff}$ range. Differential iron
abundances were calculated by subtracting to each FeI and FeI line the average
FeI abundance. FeI lines have a flat trend (the black continuos
line) while FeII lines are spread around the mean FeI abundance. This is not
surprizing since Temperature and gravity were obtained in order to have flat
trand for FeI lines and to match mean FeI and FeII abundances. However this
test shows that the precedure was applied correctly. 
The [Fe/H] trend could be due to departure from the LTE approzimation we are
using in our data analysis (the NLTE effect). In order to check this, 
for a subsample of 23 Fe I lines we obtained NLTE correction from the INSPEC
\footnote{version 1.0  (http://inspect.coolstars19.com/index.php?n=Main.HomePage)} database.
The mean NLTE correction for \#1236 is +0.06 while for \#1293$_1$ id +0.01.
This goes in the wrong direction, because the net effect would be to make the
warmer stars even more metal rich compared with the coolers. However, as shown
in Fig.~\ref{f2d} (lower panel), there is a trend of the NLTE correction with the exitacion
potential. This is not unexpected because lines with low E.P. form in the upper
atmosphere where gas density is lower and detarture from LTE is larger.
While for stars \#1293$_1$ (T$_{eff}$$\sim$4000K, red points) the NLTE correction
is very small along the entire E.P. range and can be neglected, for stars  
\#1236 (T$_{eff}$$\sim$4600K, black points) reaches $\sim$0.1 dex for 
lines with E.P.=1. For this star we performed again the spectroscopic
analysis applaying to FeI lines the NLTE correction of Fig.~\ref{f2d} (lower panel, black line).
The result was a slight change in the atmospheric parameters (T$_{eff}$=4565 K,
log(g)=1.70, and v$_{t}$=1.40 km/s) but a negligible change in metallicity
([Fe/H]$_{LTE/NLTE}$=-1.16/-1.17). This is because NLTE corrections cause a downward
change of the temperature that counterbalances the mean +0.06 dex correction
meantioned before. The conclusion of this test is that NLTE cannot be the
reason of the [Fe/H] trend.

If we look carefully to the trend however, we notice that stars between
T$_{eff}$=4100K and T$_{eff}$=4520K show a very flat slope
that deviates from the 0 value of less then 0.2 $\sigma$. So the [Fe/H] trend
is entirely due to the three warmer and to the three cooler points,
that deviate from the mean iron abundance of $\sim$1.5 $\sigma$ and $\sim$1.0
$\sigma$ respectivelly. Actually Fig.~\ref{f2a}  shows that 5 out of 21 points
deviate of 1$\sigma$ or more with respect to the mean [Fe/H] abundance, that
represents 24\% of the sample.
This matches very well with deviation that is expected statistically,
because 22\% of the points are expected to fall out of the $\sim$1$\sigma$ range.
Because of this there is no reason not to attribute the [Fe/H] trend to
statistical fluctuations. This is supported also by the error analisys (see
Tab.~\ref{t4} and below) that gives an expected r.m.s for our data of 0.06 dex that
matches very well the observed spread of 0.06 dex.
Maybe some other mechanism is at work like molecular
bands that could affect the position of the continuum for the two cooler
stars (in this case the continuum would be depressed causing an
underestimation of the EQWs and so lower abundances for iron but also for all
the other elements), but the accuracy of our measurements is not high enough
to disentagle such a mechanism from statistical noise.

As far as Silicium, Calcium, Nickel, and Barium abundance ratios are concerned all stars are spread
around the mean value and no sign of trend is present 
Only Ti appears to have a slight trend with temperature
but the range in [Ti/Fe] of our data is so small ($\sim$0.1 dex) that it is
hard to attribute this to a cause other than statistical fluctuation, at least
with the current internal accuracy of our measurements.
However, since Ti final abundance was obtained from neautral and single
ionized lines,  we plotted in Fig.~\ref{f2d} (upper and middle panels) the
differential Ti and TiII abundaces. As in the case of FeI and FeII, TiI lines follow
a flat trand while TiII lines are spread around the mean TiI abundance.
TiI and TiII are sensitive to stellar parameters but were not used to obtain them.
For this reason this check shows that the method we used to obtain T$_{eff}$ and log(g) is robust.

As a final conclusion  we can say that the methodology used to obtain chemical abundances is consistent over the
entire temperature range with possible minor trends on [Fe/H] and [Ti/Fe] with
effective temperature that anyway could
be due to statistical fluctuations and do no affect the results.

Fig.~\ref{f2b} and Fig.~\ref{f2c} show the light elements O, Na, Mg, and Al 
abundances vs. T$_{eff}$. Those are the most important elements for the
following discussion. We see that O, Na, Mg, and Al do not show any trend with T$_{eff}$
and the large spread around the mean value is due to the presence in the
cluster of stars with inhomogeneous abundances.

A detailed internal error analysis was performed using two methods.
For the first we plotted in Fig.~\ref{f3} temperature, gravity and
microturbulence of our stars as a function of the K$_{2MASS}$ magnitude (black
points).
Because the error on K$_{2MASS}$ is negligible in this plot (0.01$-$0.02
mag.), the dispersion around the parabolic fit (black lines) is entirely due
to errors on the atmospheric parameters. This dispersion gives us
$\sigma$(T$_{eff}$)=41 K, $\sigma$(log(g))=0.12, and $\sigma$(v$_{t}$)=0.04
km/s. The error on [Fe/H] due to the S/N is 0.01 dex.
As a comparison, in Fig.~\ref{f3} we plotted with blue points
our initial guess for temperature, gravity, and microturbulence as obtained from photometry together with a
parabolic fit (blue lines). As far as T$_{eff}$ is concerned, for warmer and cooler targets the match is
reasonably good while in the middle region photometric temperatures are about
100-150 K higher. The mean difference is about 60 K. We underline the fact that if we would have adopted photometric
temperature, targets in the middle region would have had a [Fe/H] value 0.15-0.20
dex larger then the others, while with the temperature scale we adopt in this
paper the [Fe/H] value of all the stars agree well within the errors (see
Tab.~\ref{t4}). On the other hand our gravities are 0.3 dex lower on average
than the photometric values while microturbulences are 0.05 km/s lower, with
the warmer stars showing a better agreement. These mismatches
could be due to the poorly known parameters of the cluster (a change in
distance modulus of +0.1 mag and in the reddening of -0.05 mag would give an
almost negligible difference in temperature and an agreement within 0.2 dex in
gravity) or to effects not taken into account by the models (i.e. 3D effects).
\\
The second method is bases on the 4 stars observed twice (\#1246, \#1293, \#1378,
and \#1380). 
We derived the atmospheric parameters independently from the
individual spectra of each star. Then we computed the differences in Teff, log(g) and vt,
between the two determinations. The distribution of atmospheric parameters differences
gives us $\sigma$(T$_{eff}$)=33 K, $\sigma$(log(g))=0.12, and $\sigma$(v$_{t}$)=0.03
km/s and a total error on [Fe/H] of 0.03 dex. This total error takes
into account all the errors on the atmospheric parameters.\\
The two methods agrees well, and we adopted $\sigma$(T$_{eff}$)=40 K,
$\sigma$(log(g))=0.12, $\sigma$(v$_{t}$)=0.04, and $\sigma$([Fe/H])=0.03 dex
as our finals errors on the atmospheric parameters.
Then we choose star \#1367 as representative of the sample,
varied its T$_{\rm eff}$, log(g), [Fe/H], and v$_{\rm t}$ according the
the atmospheric errors we just obtained, and redetermining the abundances.
Results are shown in Tab.~\ref{t4}, including the error due to the noise
of the spectra. This error was obtained for elements whose abundance was
obtained by EQWs, as the errors on the mean given by
MOOG, and for elements whose abundance was obtained by spectrum-synthesis, as
the error given by the fitting procedure. $\Delta_{\rm tot}$ is the
squared sum of the single errors, while $\Delta_{\rm tot}(obs)$ is the error
as obtained from the 4 repeated stars. $\Delta_{\rm tot}(obs)$ is equal or smaller
than $\Delta_{\rm tot}$. In Tab.~\ref{t4} for each element we report the
observed spread of the sample ($RMS_{\rm obs}$) with its error and in the final column the
significance (in units of $ \sigma$) calculated as the absolute value of the difference between 
$RMS_{\rm obs}$ and $\Delta_{\rm tot}$ divided by the error on $RMS_{\rm obs}$ (we choose $\Delta_{\rm tot}$ instead of
$\Delta_{\rm tot}(obs)$ to be conservative). This tells us if the observed dispersion $RMS_{\rm obs}$ is
intrinsic or due to observational errors. Values larger than 3$\sigma$ imply
an intrinsic dispersion in the species chemical abundance among
the cluster starst.

\begin{figure}
 \includegraphics[width=80mm]{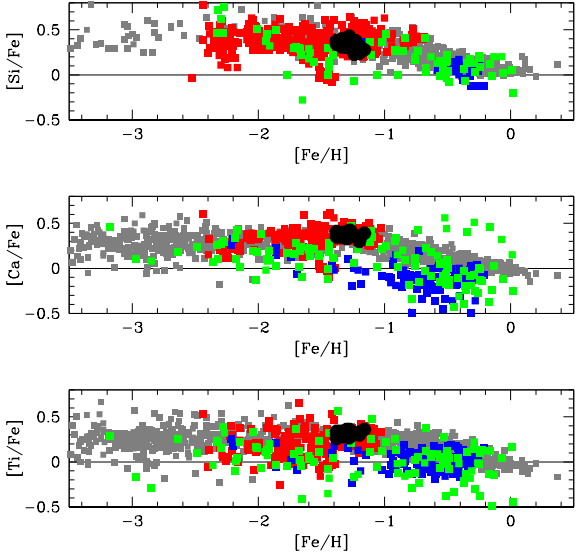}
 \caption{Star by star (black filled circles) abundances for M28 compared with a variety of Galactic and extragalactic
   environments: Galactic Globular Clusters (red filled squares); Disc and 
  Halo stars (gray filled squares); Magellanic clouds (blue filled squares);
  other dwarf and ultra-faint dwarf galaxies (Draco, Sextans, Ursa Minor, 
  Sagittarius, Bootes I and Hercules, green filled squares). See text for more details.}
 \label{f4}
\end{figure}

\section{Results}

\subsection{Iron-peak, $\alpha$ elements and heavy elements}

The mean iron content we obtained is:

$$[Fe/H]=-1.29\pm0.01$$

with a dispersion of:

$$\sigma_{[Fe/H]}=0.06\pm0.01$$

Reported errors are errors on the mean. \citet{Ha96} gives [Fe/H]=-1.32
and we can consider the agreement satisfactory.
The measured iron dispersion in Tab.~\ref{t4} well agrees with the 
dispersion due to measurement errors so we no evidence for an intrinsic Fe abundance
spread. As far as iron-peak elements are concerned, Cr and Ni are solar scaled, Mn and
Cu are sub-solar, and V, Co and Zn are super-solar. 

The $\alpha$ elements Si, Ca, and Ti are overabundant compared to the
Sun. This is a feature common to almost all Galactic GC and Halo field stars as well as to very metal-poor
stars ([Fe/H]$<$-1.5) in outer galaxies. 
Based on these elements we derive for the cluster a mean $\alpha$ element
abundance of:

\begin{center}
$[\alpha/Fe]=+0.34\pm0.01$
\end{center}

As far as heavy elements are concerned Y, Zr, Ba, and Eu are super-solar, La
is solar-scaled and Ce is sub-solar.

Figure \ref{f4} shows the star by
star (black filled circles) $\alpha$-element abundances of the cluster compared with a variety of
galactic and extra-galactic objects. We have included values from Galactic
Globular Clusters (GCs) \citep[red filled squares]{Ca09b,Ca10,Ca14a,Ca14b,Ca15a,Ca15b,Vi10,Vi11,Vi13,Mu13,Sa15}; Disc and Halo stars
\citep[gray filled squares]{Fu00,Re03,Re06,Ca04,Si04,Ba05,Fr07,Jo12,Jo14}
and extra-galactic objects such as Magellanic clouds \citep[blue filled squares]{Po08,Jo06,Mu08,Mu09},  
Draco, Sextans, Ursa Minor and Sagittarius dwarf galaxy and
the ultra-faint dwarf spheroidals Bo\"{o}tes I and Hercules
\citep[green filled squares]{Mo05,Sb07,Sh01,Is14,Ko08}.

The $\alpha$ elements in M28 follows the same trend as Galactic GCs and
are fully compatible with Halo field stars while it falls in a region 
scarcely populated by extragalactic objects.
So, according the its  $\alpha$-element content, M28
is very likely a genuine Galactic cluster.

\begin{figure}
 \includegraphics[width=80mm]{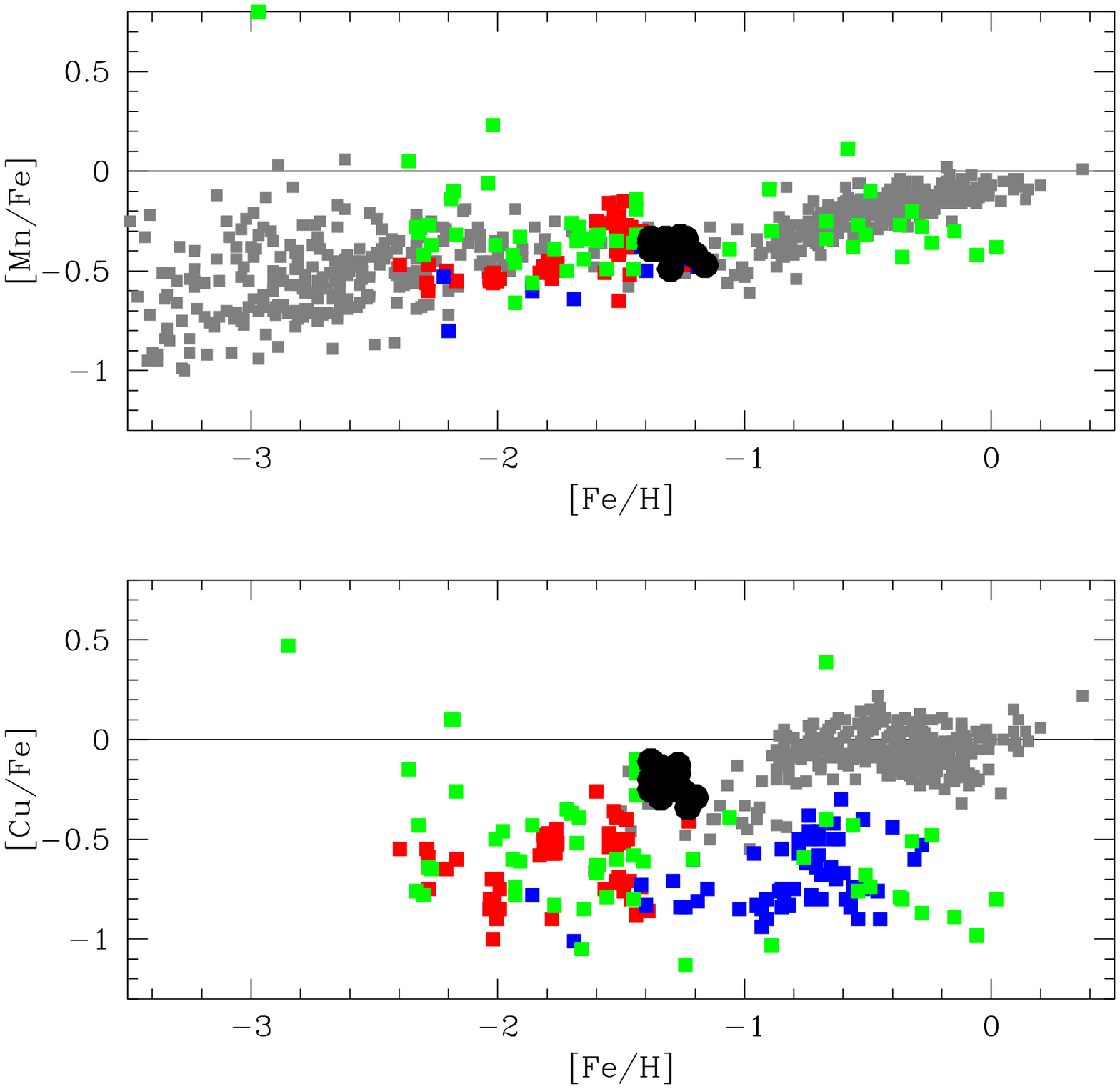}
 \caption{Star by star (black filled circles) abundances for M28 compared with a variety of Galactic and extragalactic
   environments: Galactic Globular Clusters (red filled squares); Disc and 
  Halo stars (gray filled squares); Magellanic clouds (blue filled squares);
  other dwarf and ultra-faint dwarf galaxies (Draco, Sextans, Ursa Minor, 
  Sagittarius, Bootes I and Hercules, green filled squares). See text for more details.}
 \label{f5}
\end{figure}

The chemical abundances for the iron-peak elements Mn and Cu are reported in
Fig.~\ref{f5}. Around M28 metallicity, Galactic and extragalactic environments share the
same Mn abundance, while the cluster Cu content agrees better with the
Galaxy. This further supports a Galactic origin.

Finally for all $\alpha$, iron-peak and heavy elements Tab.~\ref{t4} shows that the
observed dispersion agrees well with the measurement errors so we can rule out
any intrinsic abundance spread. We check also for possible correlations of
these elements with lights elements such Na and Al, but we did not find
evidence for significant trends.

\begin{figure}
 \includegraphics[width=80mm]{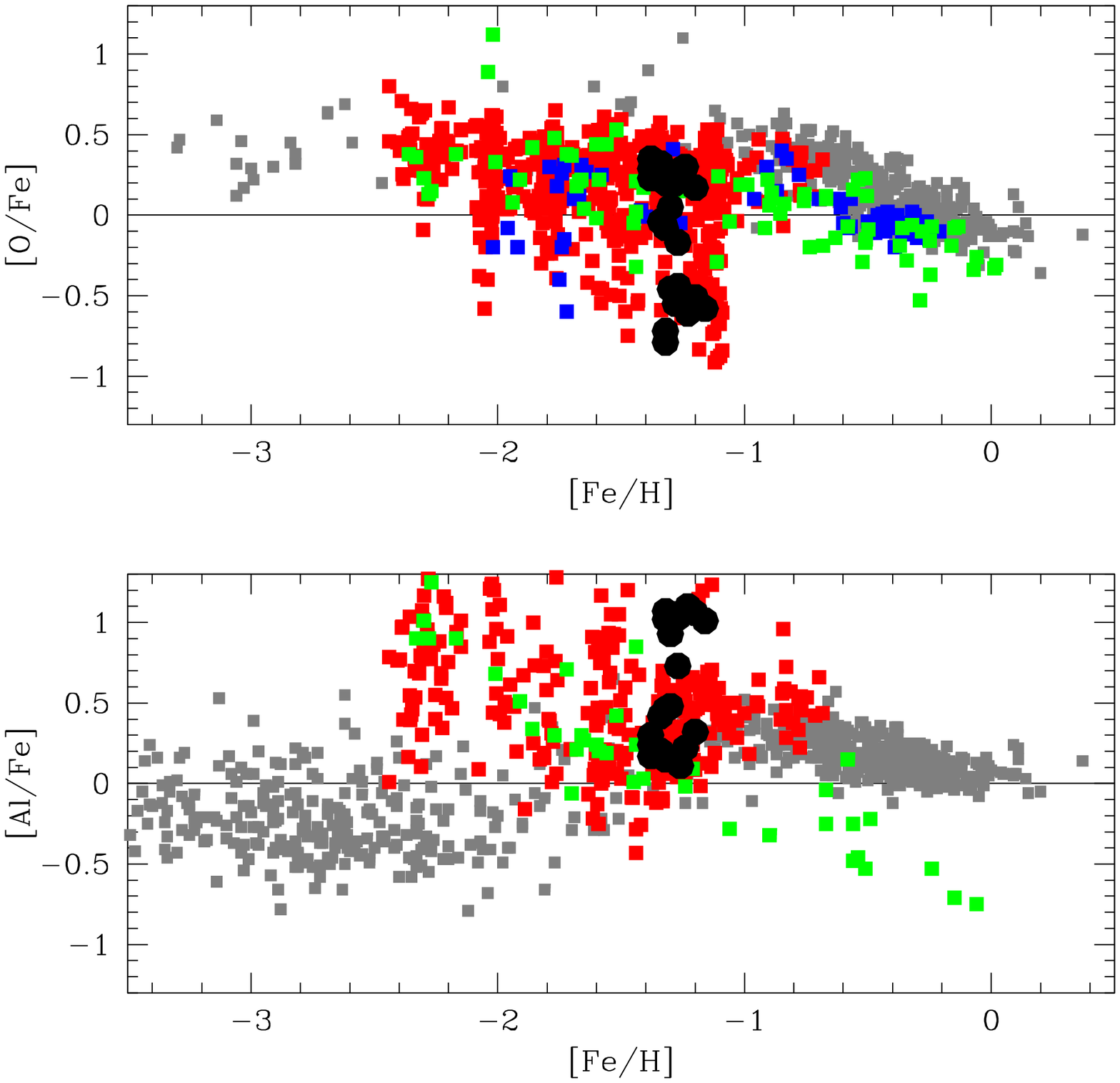}
 \caption{Star by star (black filled circles) abundances for M28 compared with a variety of Galactic and extragalactic
   environments: Galactic Globular Clusters (red filled squares); Disc and 
  Halo stars (gray filled squares); Magellanic clouds (blue filled squares);
  other dwarf and ultra-faint dwarf galaxies (Draco, Sextans, Ursa Minor, 
  Sagittarius, Bootes I and Hercules, green filled squares). See text for more details.}
 \label{f6}
\end{figure}

\begin{figure}
 \includegraphics[width=80mm]{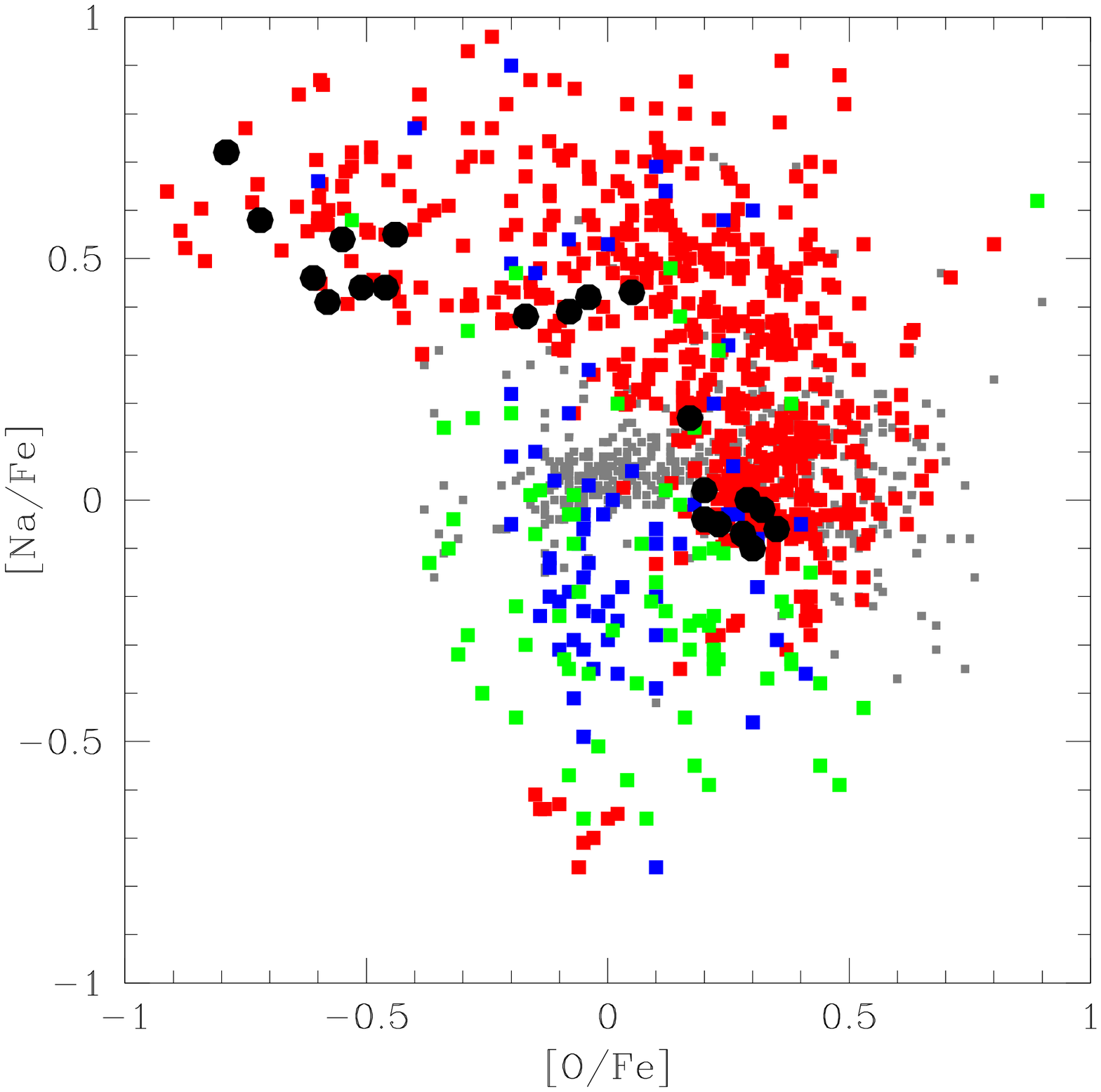}
 \caption{Na-O anticorrelation for M28 (black filled circles) compared with a
   variety of Galactic and extragalactic environments: Galactic Globular Clusters (red filled squares); Disc and 
  Halo stars (gray filled squares); Magellanic clouds (blue filled squares);
  other dwarf and ultra-faint dwarf galaxies (Draco, Sextans, Ursa Minor, 
  Sagittarius, Bootes I and Hercules, green filled squares). See text for more
   details.}
 \label{f7}
\end{figure}

\begin{figure}
\centering
\includegraphics[width=8cm]{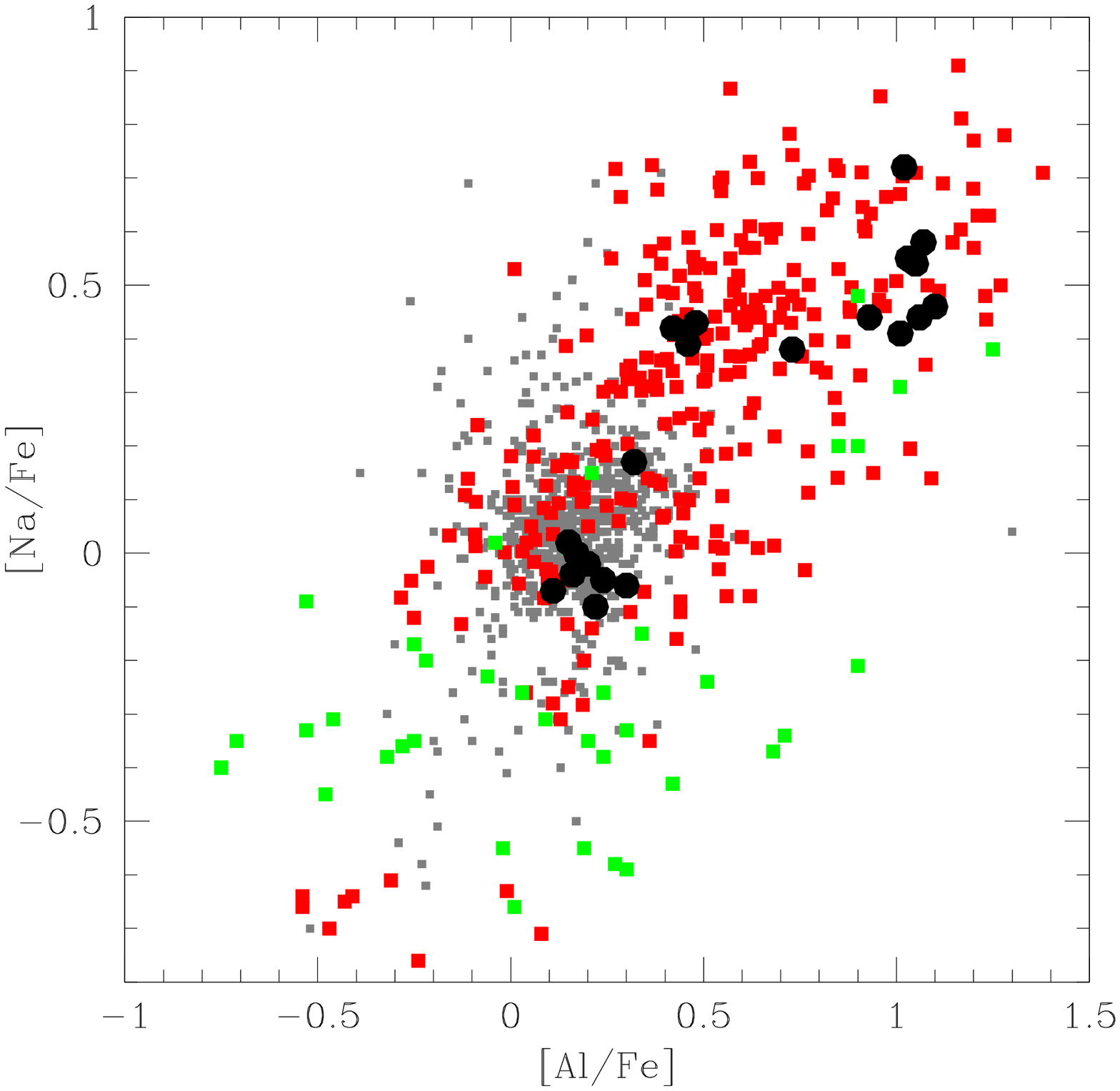}
\caption{Na-Al correlation for M28 (black filled circles) compared with a
   variety of Galactic and extragalactic environments: Galactic Globular Clusters (red filled squares); Disc and 
  Halo stars (gray filled squares); Magellanic clouds (blue filled squares);
  other dwarf and ultra-faint dwarf galaxies (Draco, Sextans, Ursa Minor, 
  Sagittarius, Bootes I and Hercules, green filled squares). See text for more
   details.}
\label{f8}
\end{figure}

\begin{figure}
\centering
\includegraphics[width=8cm]{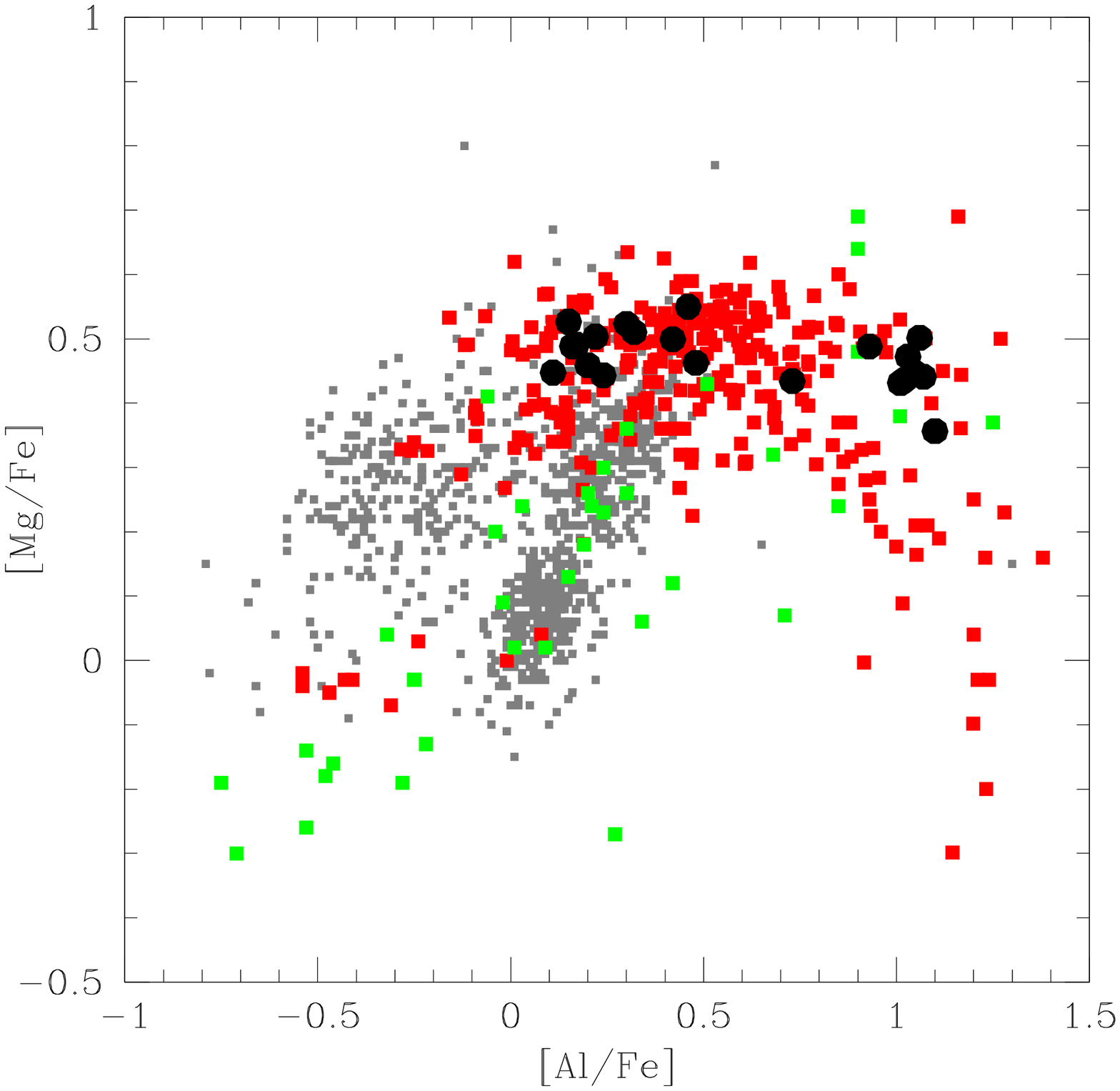}
\caption{Mg-Al anticorrelation for M28 (black filled circles) compared with a
   variety of Galactic and extragalactic environments: Galactic Globular Clusters (red filled squares); Disc and 
  Halo stars (gray filled squares); Magellanic clouds (blue filled squares);
  other dwarf and ultra-faint dwarf galaxies (Draco, Sextans, Ursa Minor, 
  Sagittarius, Bootes I and Hercules, green filled squares). See text for more
   details.}
\label{f9}
\end{figure}

\subsection{Light elements}

\begin{figure*}
\centering
\includegraphics[width=16cm]{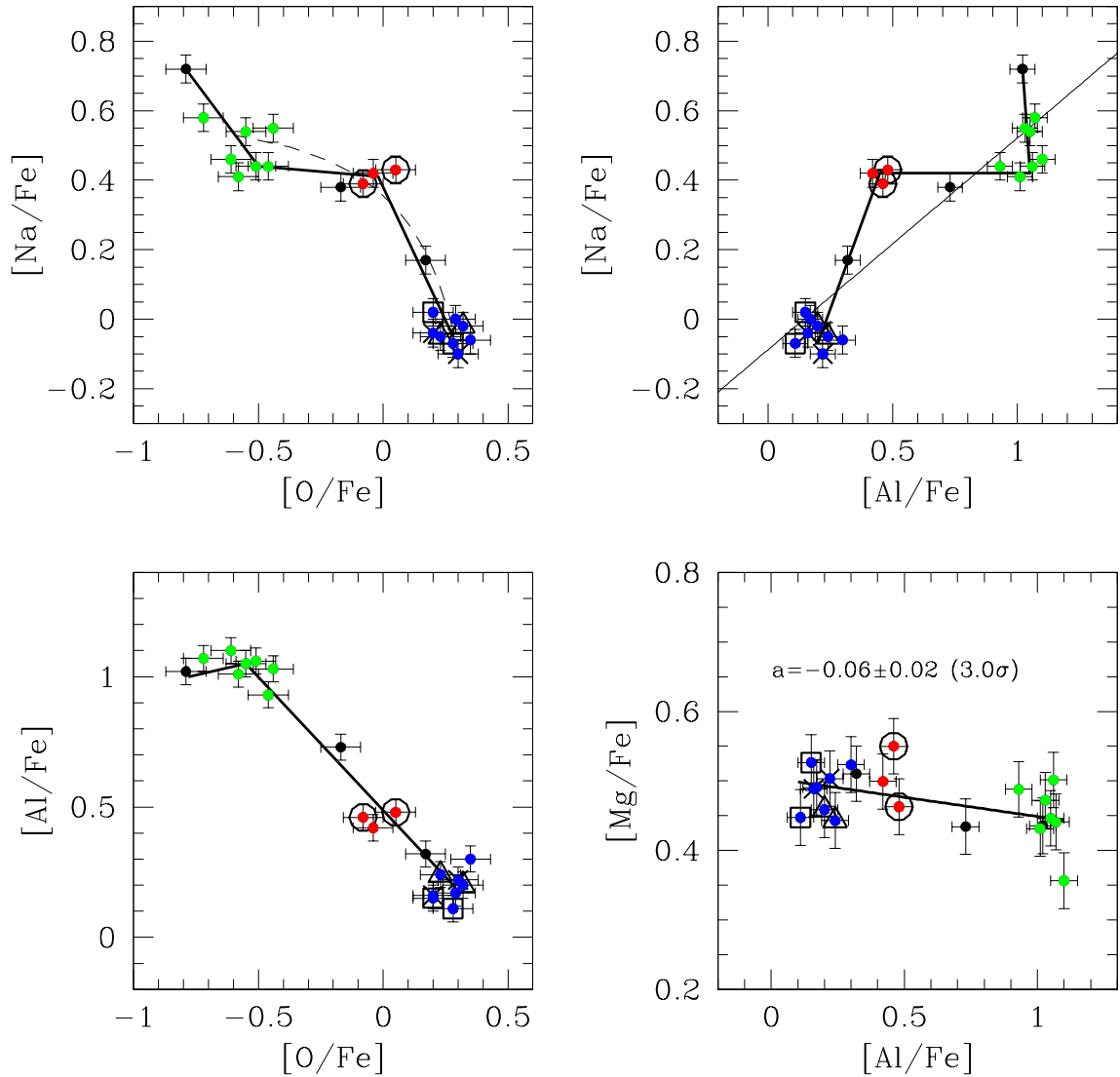}
\caption{Na-O anticorrelation (upper left panel), Na-Al correlation (upper
  right panel), Al-O anticorrelation (lower left panel), and Mg-Al
  anticorrelation (lower right panel) for M28. The 4 stars observed twice
    are indicated with 4 different black open symbols. See text for more
  details.}
\label{f10}
\end{figure*}

Light elements O, Na, and Al have an observed spread that well exceeds
the observational uncertainties (see Tab.~\ref{t4}). The only exception is Mg that
seems to be homogeneous within the errors.\\
In Fig.~\ref{f6} and Fig.~\ref{f7} we compare the O and Al abundances of the targets with different
environments. The content of the two elements of the M28 first generation
stars (the targets with [O/Fe]$\sim$0.3, [Na/Fe]$\sim$0.0, [Al/Fe]$\sim$0.2)
well matches the mean O and Al abundances of
the Milky Way Halo and the mean O and Al content of the other GC first
generation stars. It matches also the O and Al content of extragalactic
environments. On the other hand the cluster shows a strong depletion in O
and a strong Al enhancement as far as the subsequent generation targets
are concerned (the targets with [O/Fe]$<$0.3, [Na/Fe]$>$0.1,
[Al/fe]$>$0.2). Actually M28 is one of the clusters that shows the strongest O
depletion (down to [O/Fe]$\sim$-0.8) and the strongest Al enhancement (up
to [Al/Fe]$\sim$+1.1) together with few other GCs in the Milky Way
(e.g. NGC~2808, NGC~3201, and NGC~6752) and in the Large Magellanic Cloud (e.g. NGC~1718).\\
Fig.~\ref{f7} shows the Na-O anticorrelation for M28. First generation stars
have a Na content that is lower than the bulk of Milky Way field objects.
The second generation stars follow the Galactic GC trend along the O-poor edge (left edge). This
is because M28 has one of the lowest initial O contents ([O/Fe]$\sim$0.3) with respect to the
other GCs. This relatively low O content is maintained when subsequent generations
stars are formed. The cluster population ends up with two stars (\#1291 and
\#1295) that are among the most O-poor star in the GC sample we have.
This means that M28 suffered one of the most extensive nuclear processing.
Fig.~\ref{f7} shows also the star formation process was
not continuous, but it happened with different bursts because our targets
aggregate in at least three different clumps.\\
Fig.~\ref{f8} and Fig.~\ref{f9} show the Na-Al and Mg-Al correlations. The
cluster follows the typical correlations of the GCs, with Al-rich stars
([Al/Fe]$>$0.8) that appear to be slightly Mg-poorer that the others on
average. However the most surprising result is that in the Na-Al correlation
stars appear to be clearly aggregated in 3 clumps and that the
abundance evolution is not linear.\\
In order to investigate in more detail this behavior we plot in
Fig.~\ref{f10} light-element correlations or anticorrelations. Errorbars are from
Table~\ref{t4}(7th column). We analyze this figure starting from the Na-Al
plot (upper right panel). As said before, it is clear that stars do not
follow a linear trend (the continuos thin line) but very likely a {\it segmented} path that we indicated with
a continuous thick line. Stars below [Al/Fe]=0.5 increment their Na and Al
content linearly. After that stars appear to have a constant Na content
([Na/Fe]$\sim$0.4) while Al increase up to [Al/Fe]$\sim$+1.0. Finally stars
evolve to higher Na abundances (up to [Na/Fe]$\sim$0.7) maintaining Al
constant. 
In order to support this statement we calculated the distance of the
points from the linear trend. If we decompose this distance into a vector
along the X axis, that corresponds to the deviation of the [Al/Fe] abundance of
the point, and into a vector along the Y axis, that corresponds to the deviation
of the [Na/Fe] abundance of the point, we obtain the deviations of the [Na/Fe]
and [Al/Fe] abundances of our data with respect the linear trend. 
The r.m.s of these deviations would imply an error on our [Na/Fe] values of
0.09$\pm$0.01 dex and an error on our [Al/Fe] values of 0.06$\pm$0.01
dex. If we assume that the error on $\Delta_{\rm tot}$(Na) 
is also 0.01 as Tab.~\ref{t4} suggests, we find that the value for [Na/Fe] exceeds more than
3$\sigma$ the error estimation of Tab.~\ref{t4} according to the equation:

$$Significance=(0.09-0.04)/(\sqrt(0.01^{2}+0.01^{2}))\ \sigma=3.5\ \sigma$$

and that for [Al/Fe] is larger too. So we conclude that a linear trend is a
very poor approximation of the data. 

In this evolutionary path our targets are not distributed uniformly
but aggregate in clumps. There are at least 3 of them at
1) [Al/Fe]$\sim$0.2, [Na/Fe]$\sim$-0.1 (blue points), [Al/Fe]$\sim$0.4,
[Na/Fe]$\sim$0.4 (red points), and 3) [Al/Fe]$\sim$1.0, [Na/Fe]$\sim$0.5
(green points) with some interlooper. The
third group has a spread much larger than the other two (up to 0.2 dex), and
could be composed of two further sub-populations. However more data are
required to constrain better the number of sub-populations in M28.
We underline the fact that it is not unusual for a GC to show discrete sub-populations. As an
example \citet{Mi15} found that NGC~2808 is formed by 5 or more discrete
sub-populations.

In the upper left panel of Fig.~\ref{f10} we report the Na-O anticorrelation.
We fitted the \citet{Ca09a} dilution model (dashed black line). 
According to this fit the cluster is composed
by a first generation of stars with [Na/Fe]$\sim$-0.10 and [O/Fe]$\sim$+0.30 and
subsequent generations ends up with [Na/Fe]$\sim$+0.55 and [O/Fe]$\sim$-0.6.
However, while this model fits well the blue and red
sub-populations, it fails for the green. In particular the most O-poor stars
show a spread in Na that is not reproduced at all by the dilution model. Based
in what we have found above, we 
fitted a segmented anticorrelation represented by the continuous black line. 
Stars above [O/Fe]=-0.3 increment their Na and decrement they O
contents linearly. After that stars appear to have a constant Na content
([Na/Fe]$\sim$0.4) while O decreases down to [O/Fe]$\sim$-0.5. Finally stars
evolve to higher Na abundances (up to [Na/Fe]$\sim$0.7) and lower O abundances
(down to [O/Fe]$\sim$-0.7) linearly. \\
In the lower panel on the left we report the Al-O anticorrelation. At odd
with the previous plots, the anticorrelation is linear down to
[O/Fe]$\sim$-0.5 with aluminum that increases while oxygen decreases. After
that the correlation seems to twist with very O-poor stars that decrease their
aluminum content.\\
Finally in the lower panel on the right we plot [Mg/Fe] vs. [Al/Fe]. The error analysis did not
show evidence for a intrinsic spread of Mg for our stars. On the other end
here we see that Al-rich stars are also Mg-poor and the trend has a slope with
a significance at the level of 3$\sigma$. This is not unexpected since 
aluminum is produced by the Mg-Al chain at the expenses of Mg \citep{Ve11} in
intermediate-mass AGB stars, that are one of the main candidates responsible for the
multiple-population phenomenon and that are the only candidates able to activate this chain since
they reach high enough temperatures ($\sim$75 million K) during the hot-bottom-burning phase \citep{Da16}. 

\section{Summary}

In this paper we present the first detailed chemical abundances of 21 elements
in 17 red giant radial velocity and metallicity members of M28 observed using the high resolution UVES
spectrograph, mounted at the VLT-UT2 telescope. M28 is a metal poor GC ([Fe/H]$\sim$-1.3) and
must be very old ($\sim$13 Gyrs) because of the HB morphology that completely
lacks a red branch. Chemical abundances have been computed using plane-parallel atmospheric-models
and LTE approximation. Equivalent width method has been used when possible. Otherwise we applied
the spectrum-synthesis method. We obtained the following results:

\begin{itemize}
 \item We found a mean metallicity of [Fe/H]=-1.29$\pm$0.01, that well agree
       with previous literature data. As far as other iron-peak elements are
       concerned, Cr and Ni are solar-scaled, Mn and Cu are sub-solar, and V,
       Co, and Zn are super-solar.
 \item M28 has the typical $\alpha$-enhancement of the Halo with [$\alpha$/Fe]=+0.34$\pm$0.01. 
       Heavy elements Y, Zr, Ba, and Eu are super-solar, La is solar scaled and Ce
       is sub-solar. We did not find any intrinsic spread in any of these
       elements, and no correlation with light elements.
 \item M28 shows the typical Na-O anticorrelation common to almost all
       the other GCs. The cluster also show a Na-Al correlation and a Mg-Al
       anticorrelation. It is one of the clusters that shows the strongest O
       depletion and Na and Al enhancement among all the GCs studied up to now.
       The presence of a Mg-Al anticorrelation points toward intermediate-mass AGB stars
       as the main polluters responsible for the multiple-population
       phenomenon in this cluster.
 \item The most interesting results however concerns the shape and
       discreteness of the Na-Al and Na-O relations. Both appears to be not
       linear or parabolic, but {\it segmented}, and stars are not distributed
       continuously, but seems to form at least 3 different sub-populations.
       This is not totally new since also the GC NGC~2808 shows the presence
       of at least 5 distinct sub-populations. A larger sample of
       data is required to confirm if more sub-populations are present.
\end{itemize}

\section*{Acknowledgments}

SV and CMB acknowledge the support provided by Fondecyt Regular 1130721 and
1150060, respectively. FM gratefully acknowledges the support provided by
Fondecyt for project 3140177. LM acknowledges support from {\it "Proyecto Interno"} of the
Universidad Andres Bello.

\bsp

\label{lastpage}

\end{document}